\documentclass[a4paper,12pt]{article}
\usepackage{mathrsfs}
\usepackage{graphicx} 
\usepackage{epstopdf}
\usepackage{subfigure}
\usepackage{color,txfonts}
\definecolor{blue0}{rgb}{0,0,0.6}
\usepackage[colorlinks,linkcolor=blue0,anchorcolor=blue0,citecolor=blue0,urlcolor=blue0]{hyperref}

\begin{document}

\hoffset = -1truecm \voffset = -2truecm \baselineskip = 10 mm

\title{\bf Gluon condensation: from nucleon to Galactic center}

\author{Wei Zhu$^a$\footnote{Corresponding author, E-mail: wzhu@phy.ecnu.edu.cn},
Zi-Qing Xia$^b$,  Yu-Chen Tang$^b$ and Lei Feng$^b$ \\\\
         \normalsize $^a$Department of Physics, East China Normal University,\\
       \normalsize Shanghai 200241, P.R. China\\
       \normalsize $^b$Key Laboratory of Dark Matter and Space Astronomy, Purple Mountain Observatory,\\
         \normalsize Chinese Academy of Sciences, Nanjing 210008, P.R. China}

\date{}

\newpage

\maketitle

\vskip 3truecm

\begin{abstract}

The Galactic Center Excess (GCE), one of the most remarkable discoveries by Fermi-LAT, has prompted extensive exploration over the past decade, often attributed to dark matter or millisecond pulsars. This work proposes a novel interpretation on the origin of the GCE, focusing on the observed spectral shape. Protons are accelerated at the Galactic center and collide with the neutron cluster on the surface of the non-rotating neutron stars. Due to the gluon condensation in nucleons, these collisions produce a large number of mesons, which have reached to the saturation state and subsequently generate the broken power law in the gamma ray spectra. We explained the spectral shape of GCE using the gluon condensation and an assumption of existing the non-rotating neutron stars at the Galactic center. This example of the gluon condensation mechanism not only expands the applications of the hadronic scenario in the cosmic gamma ray spectra but also provides a new evidence of the gluon condensation.

\end{abstract}

{\bf keywords} Galactic center excess; Gluon condensation; Broken power law; Non-rotating neutron star

\newpage

\vskip 1truecm
\section{Introduction}

    The Galactic center is a hub of high-density matter, comprised of neutron stars, black holes and enigmatic
dark matter. This environment presents an exceptional opportunity for astronomers and physicists to study astrophysical objects in a highly compacted region. The Fermi Large Area Telescope (Fermi-LAT) has detected an excess of gamma rays originating from the Galactic center [1-3], known as the Galactic Center Excess (GCE). The GCE is a highly intriguing mystery in astrophysics, and deciphering the physical processes behind this phenomenon presents a challenge [4,5].

    Early investigations have indicated that one of the possible GCE sources origins from the annihilation of weakly
interacting massive particles (WIMPs), which is one of the leading dark matter candidates. Meanwhile, the GCE spectra bear resemble to the {\rm GeV} gamma-ray spectra observed in pulsars, despite the identity of their progenitor pulsars remaining unclear.

    A difficult problem is that the GCE spectra are significantly influenced by the Galactic diffuse emission [6-13].
It is critical to eliminate these contributions from the GCE spectra.
Recently, Dinsmore and Slatyer utilized a generalized Navarro-Frenk-White squared spatial template to model pulsar distributions in the Galactic center region and extracted the corresponding GCE flux from nine previous GCE energy spectra [14].
The Region of Interest (ROI) used by them was a square region selected by Galactic latitudes $\left| b\right| \textless 20^{\circ}$ and Galactic longitudes $\left| l\right| \textless 20^{\circ} $, with a mask of the Galactic plane with $\left| b\right| \textless 2^{\circ}$. The nine inferred GCE spectra in [14] show some differences from each other due to different fitting approaches, ROI choices, signal models and background templates. However, unexpectedly, all nine GCE spectra exhibit strong evidence of the typical broken power law (BPL) shape.
This consistent presence of the BPL shape across the diverse GCE spectra, despite the complexities introduced by different analyses and modeling approaches, introduces a tantalizing puzzle that beckons for a deeper comprehension of the underlying physical processes driving the GCE.

\begin{figure}
  \begin{center}
   \includegraphics[width=0.8\textwidth]{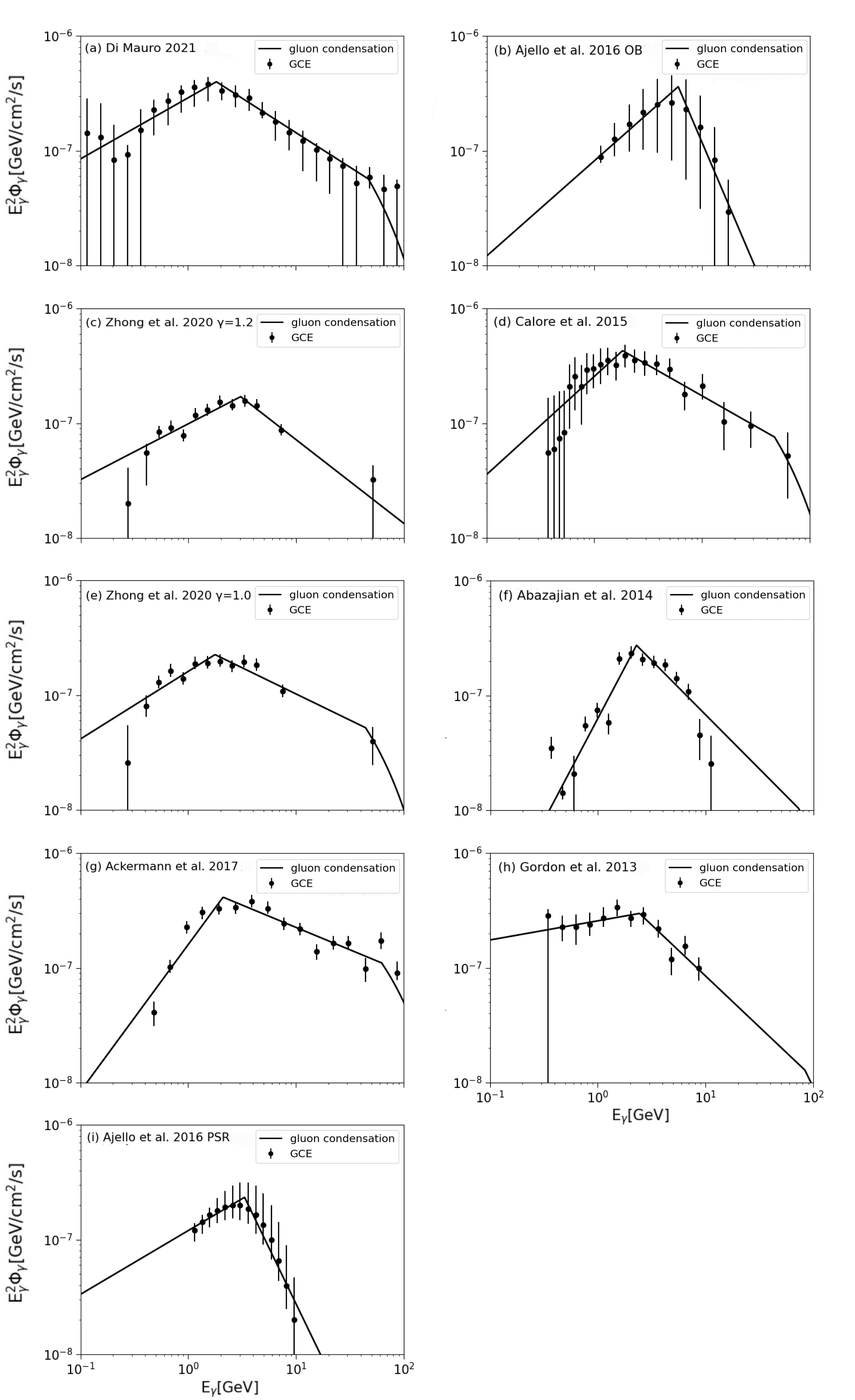}
   \caption{The GCE-spectra predicted by the gluon condensation mechanism. The data are taken from [6-14]. The results show
   the broken power law.
}\label{fig:1}
  \end{center}
\end{figure}

    This article attempts to attribute the observed GCE spectra in Fig.1 to the gluon condensation effect in nucleon (proton and neutron).
As we know, high energy nucleon collisions are general processes in the Galactic center. The hadronic scenario of gamma-ray emission is such $pp\rightarrow \pi^0\rightarrow 2\gamma$, where a bump of energy spectrum near $\sim 1~{\rm GeV}$ (i.e. the "$\pi$-bump")
origins from a peak in $\pi^0$ decay [15], while the gluon distributions in nucleon dominate the $\pi$-yield $N_{\pi}$ at high energy [16]. The evolution equations in quantum chromodynamics (QCD) predict that the gluon distributions may smoothly tend towards an equilibrium state between the splitting and fusion of gluons, known as the Color Glass Condensation (CGC) [17]. Following them, one of us (W.Z.) first proposed that the CGC distributions will continually evolve and result in a chaotic solution [18-21]. This leads to dramatic chaotic oscillations that cause strong shadowing and antishadowing effects, forcing a large number of soft gluons to condense into a state at a critical momentum $(x_c, k_c)$, where $x_c$ is a fraction of the proton's longitudinal momentum carried by the condensed gluon and $k_c$ is their transverse momentum. This is the gluon condensation (Fig.2). We will illustrate in Sec.2 that the sharp peak in the gluon distributions can significantly enhance
the cross section of the nucleon collisions and results a new bump in the gamma ray spectra, which is stronger than the ``$\pi$-bump" and has a typical BPL form with an exponential suppression factor.

\begin{figure}
  \begin{center}
   \includegraphics[width=0.8\textwidth]{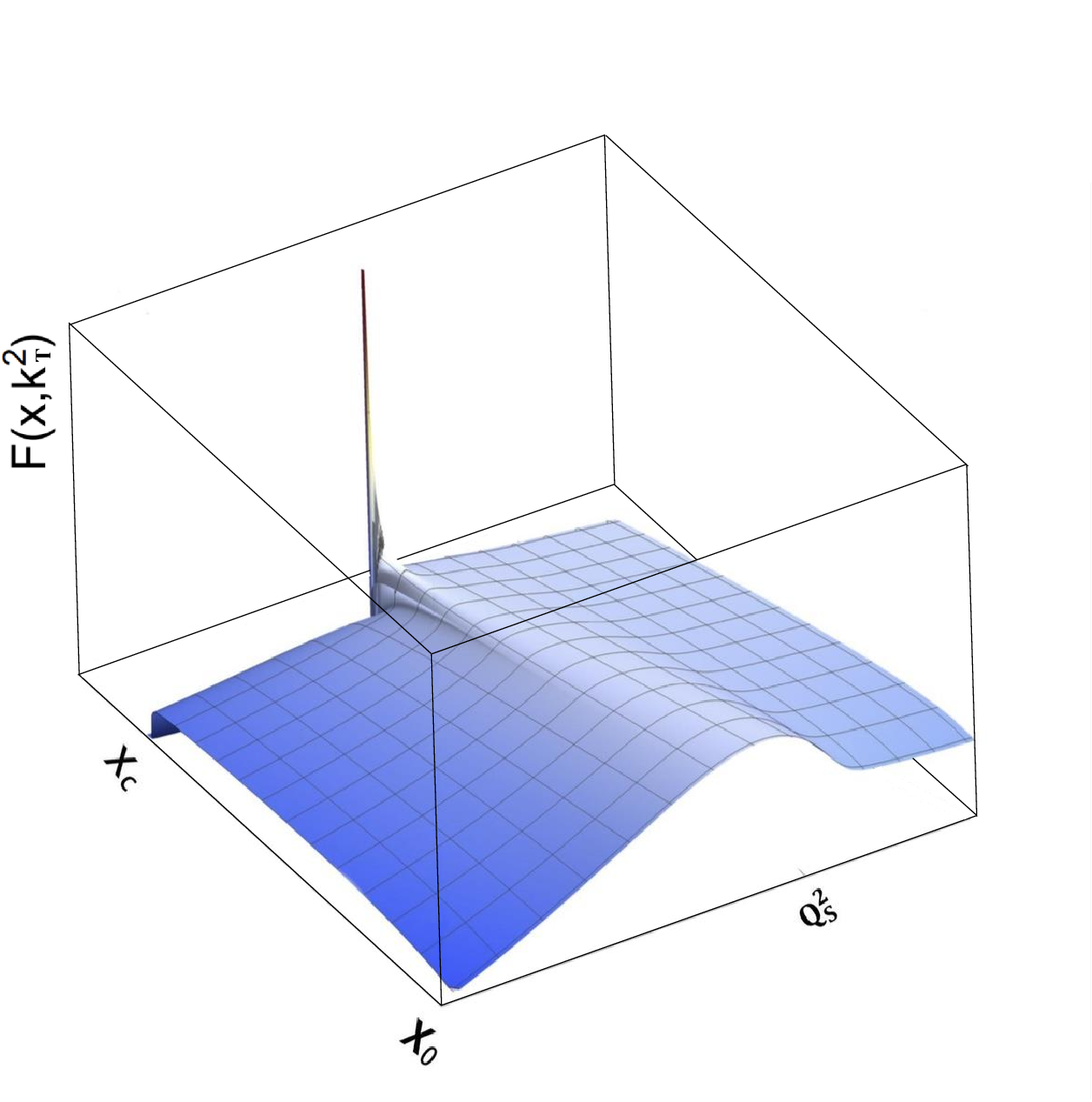}
   \caption{A schematic solution of a QCD evolution equation [21], which shows the evolution of
        transverse momentum dependent distribution $F(x,k^2_T)$ of gluons in nucleon
        from the CGC at $x_0$ to the gluon condensation at $x_c$. Note that all gluons with $x<x_c$ are stacked at $(x_c,k_c)$
}\label{fig:2}
  \end{center}
\end{figure}

      Using the gluon condensation mechanism, we will explain the GCE spectra in Sec.3.
We find that not only in the spectral shape but also in the range of model parameters, the GCE and the millisecond pulsar spectra are similar.
It implies that all of them originate from the same emission mechanism, i.e., the collision of high energy proton with the big neutron cluster $A^*\gg 300$ ($A^*$ is the number of neutrons) on the surface of a neutron star.

    However, there are not enough observed pulsar spectra to fully cover the GCE curve.
Moreover, even if enough millisecond pulsars are discovered in the Galactic center, it is difficult to synthesize a single BPL spectrum from a set of BPL distributions with diverse spectral parameters. One solution to this problem is to assume the existence of a large population of non-rotating (including slower rotating) neutron stars at the Galactic center. Although the non-rotating neutron stars are one of the theoretical topics in nuclear physics and astronomy, their observational data are extremely rare since the lack of their characteristic radiation spectra. The gluon condensation mechanism of the GCE may arouse interest in the non-rotating neutron stars.

    The information about the GCE consists of spectral shape and its spatial morphology, the latter being related to the spatial distribution of the non-rotating stars in the Galactic center.
The explanation of GCE requires confirmation not only in terms of its spectrum but also in its spatial distribution.
Unfortunately, there is no observation on the non-rotating neutron stars due to lack of detectable effects except the gluon condensation. We simply assume that the spatial distribution of the non-rotating neutron stars are similar to the millisecond pulsars, as given in Sec.4. Finally, in the last section, we give discussions and a brief summary.

\section{The gluon condensation mechanism}

   In the hadronic scenario, cosmic gamma rays can be generated through the process $pp\rightarrow \pi^0\rightarrow 2\gamma$
and they are calculated by [22]

$$\Phi_{\gamma}(E_{\gamma})=C_{\gamma}\left(\frac{E_{\gamma}}{{\rm GeV}}\right)^{-\beta_{\gamma}}
\int_{E_{\pi}^{min}}^{E_{\pi}^{cut}}dE_{\pi}
\left(\frac{E_p}{{\rm GeV}}\right)^{-\beta_p}$$
$$\times N_{\pi}(E_p,E_{\pi})
\frac{d\omega_{\pi-\gamma}(E_{\pi},E_{\gamma})}{dE_{\gamma}},
\eqno(1)$$ where the spectral index $\beta_{\gamma}$ includes the photon loss due to the medium absorption of pions.
The accelerated protons obey a power law $N_p\sim E_p^{-\beta_p}$ in the source.
$C_{\gamma}$ incorporates the kinematic factor and the flux dimension.

    On the other hand, according to the hadronic collisions model, about the half energies of parent protons are taken away
by the valence quarks, which form the leading particles, and the remaining energies are transformed into the
secondary hadrons (mainly pions) in the central region through gluon interactions.
The cross section of an inclusive particle
production in high-energy proton-proton collision is dominated by
the production of gluon mini-jet using the unintegrated gluon
distribution via [23,24]

$$\frac{dN_g}{dk_T^2dy}=\frac{64N_c}{(N^2_c-1)k_T^2}\int_{q_{T,min}}^{q_{T,max}}q_T d
q_T\int_0^{2\pi}
d\phi\alpha_s(\Omega)$$
$$\frac{F(x_1,\frac{1}{4}(k_T+q_T)^2)F(x_2,\frac{1}{4}
(k_T-q_T)^2)}{(k_T+q_T)^2(k_T-q_T)^2}, \eqno(2)$$where $k_T$ and $q_T$ are the transverse momenta,
$\Omega=Max\{k_T^2,(k_T+q_)^2/4, (k_T-q_T)^2/4\}$ and the longitudinal
momentum fractions of interacting gluons are fixed by kinematics
$x_{1,2}=k_T e^{\pm y}/\sqrt{s}$.

    A key step is how to calculate $N_g\rightarrow N_{\pi}$ at $pp\rightarrow \pi$, which involves unknown non-perturbative QCD effects.
Fortunately, a steep and high peak in the gluon distributions can determine $N_{\pi}(E_{\pi},E_p)$
and lead to a typical BPL spectrum as shown in Fig.1.  Let us to repeat this simple derivation in [25-27].

     Experiments show that $N_{\pi}$ increases as the collision energy increases, because more gluons participate in the
process of making new particles. So when a significant amount of condensed gluons at the threshold $x_c$
suddenly participate in the $pp$-collisions, it inevitably results in a large number of secondary meson production dramatically. However, since meson has mass, their yield $N_{\pi}$ is limited. As a limiting case, we imagine that $N_{\pi}$ reaches its maximum value due to the gluon condensation effect, i.e., almost all available kinetic energies of collisions at the center-of-mass (C.M.) frame are used to create pions. Of course, the validity of this approximation will be tested by the following observed data. By taking this limit, we can avoid the complicated hadronization mechanism and write energy conservation

$$E_p+m_p=\tilde{m}_p\gamma_1+\tilde{m}_p\gamma_2+N_{\pi}m_{\pi}\gamma,  \eqno(3)$$where $\tilde{m}_p$ marks the leading particle and
$\gamma_i$ is the Lorentz factor. The square of relativistic invariant total energy $s=(p_1+p_2)^2$ in the laboratory (Lab) frame and the C.M. frame are

$$2m_p^2+2E_pm_p=(2\tilde{E}^*_p+N_{\pi}m_p)^2. \eqno(4)$$Using

$$2\tilde{E}_p^*\equiv \left(\frac{1}{k}-1\right)N_{\pi}m_{\pi}, \eqno(5)$$ $k\simeq 1/2$ is an inelastic factor and it is irrelevant to
the system, we have

$$\tilde{m}_p\gamma_1+\tilde{m}_p\gamma_2=\left(\frac{1}{k}-1\right)N_{\pi}m_{\pi}\gamma. \eqno(6)$$

    From Eqs.(3) and (4) we have the power law for $N_{|pi}$

$$\ln N_{\pi}=0.5\ln (E_p/{\rm GeV})+a, ~~\ln N_{\pi}=\ln (E_{\pi}/{\rm GeV})+b,  \eqno(7)$$
$$~~ with~E_{\pi}
\in [E_{\pi}^{GC},E_{\pi}^{cut}], $$where $a\equiv 0.5\ln (2m_p/{\rm GeV})-\ln (m_{\pi}/{\rm GeV})+\ln k$
and $b\equiv \ln (2m_p/{\rm GeV})-2\ln (m_{\pi}/{\rm GeV})+\ln k$.

    Submitting Eq. (7) into Eq. (1), one can analytically obtain (here we neglect the simple integral calculations)

$$E_{\gamma}^2\Phi^{GC}_{\gamma}(E_{\gamma})\simeq\left\{
\begin{array}{ll}
\frac{2e^bC_{\gamma}}{2\beta_p-1}(E_{\pi}^{GC})^3\left(\frac{E_{\gamma}}{E_{\pi}^{GC}}\right)^{-\beta_{\gamma}+2} \\ {\rm ~~~~~~~~~~~~~~~~~~~~~~~~if~}E_{\gamma}\leq E_{\pi}^{GC},\\\\
\frac{2e^bC_{\gamma}}{2\beta_p-1}(E_{\pi}^{GC})^3\left(\frac{E_{\gamma}}{E_{\pi}^{GC}}\right)^{-\beta_{\gamma}-2\beta_p+3}
\\ {\rm~~~~~~~~~~~~~~~~~~~~~~~~ if~} E_{\pi}^{GC}<E_{\gamma}<E_{\pi}^{cut},\\\\
\frac{2e^bC_{\gamma}}{2\beta_p-1}(E_{\pi}^{GC})^3\left(\frac{E_{\gamma}}{E_{\pi}^{GC}}\right)^{-\beta_{\gamma}-2\beta_p+3}
\exp\left(-\frac{E_{\gamma}}{E_{\pi}^{cut}}+1\right).
\\ {\rm~~~~~~~~~~~~~~~~~~~~~~~~ if~} E_{\gamma}\geq E_{\pi}^{cut},
\end{array} \right. .\eqno(8)$$This is the gluon condensation spectrum. A phenomenological exponential cut factor in Eq. (8) describes
the fast suppression of the energy spectrum at $E_{\gamma}>E_{\pi}^{cut}$. The reason is that the gluons at $x<x_c$
have been condensed at $x_c$ and where no gluons to participate the $pA$ interaction. Thus, $E_{\gamma}$ has a upper limit
$E_{\pi}^{cut}$ in Eq.(1).

    Additionally, the two energy scales, $E_{\pi}^{GC}$ and $E_{\pi}^{cut}$ have the following relation.

$$E_{\pi}^{cut}=\beta e^{b-a}\sqrt{\frac{2m_p}{k_c^2}}\left(E_\pi^{GC}\right)^2,\eqno(9)$$where $\beta=1$ if $E_{\pi}^{GC}\ll 100~{\rm GeV}$.
Note that all energies take the {\rm GeV}-unit and

$$E_p=\frac{2m_p}{m^2_{\pi}}E_{\pi}^2 \eqno (10)$$gives the energy of incident proton
corresponding to $E_{\pi}$ in the Lab frame.

    We emphasize that the BPL in Eq.(8) is the analytic solution of the gluon condensation mechanism, rather than
the mathematical parameterized formula in the literature. The former has its QCD foundation and a clear physical picture,
while the latter is generally used in the data descriptions. Equation (8)
has been used to study the sub-{\rm TeV} gamma-ray spectra of the active Galactic nuclei (AGNs) and the very high energy (VHE) gamma ray spectra of pulsars, especially, the latter shows a similar structure with the GCE.

\section{The GCE spectra in the gluon condensation mechanism}

    We use the gluon condensation spectrum  Eq. (8) to fit all nine examples in Fig.1, where the data are extracted by work [6-14].
Note that $E_{\pi}^{cut}$ in Figs.1(b,c,f,h,i) are beyond the coordinate range according to Eq.(9).
The parameters are listed in Tab.1. It is interesting to note that these parameters
are broadly consistent with the fitting results of the same gluon condensation mechanism in the {\rm GeV} gamma ray spectra of pulsars in [29]. It seems that the GCE and the {\rm GeV} gamma-ray spectra of pulsars may have the same origin.

      The gluon condensation threshold $E_{\pi}^{GC}$ is the target-dependent.
According to the uncertainty relation, when a nucleus participates in collisions, the gluons with small $x$ from different nucleons
can fusion and enhancing the nonlinear effect.
Due to kinematic constraints, when protons collide, not all gluons are excited to participate in the interaction. The higher the proton collision energy, the smaller $x$ of the excited gluons. Therefore, a very high collision energy is required for the process of producing mesons by a large number of gluons condensing at a small critical $x_c$.
The study of QCD evolution equations shows that the larger $A$, the stronger the nonlinear effect and the larger the critical $x_c$ for generating the lower condensation threshold $E_{\pi}^{GC}$.
Specifically, the numerical computations show that the values of $E_{\pi}^{GC}$ are taken from $0.1~{\rm TeV}\sim 20~{\rm TeV}$ in the $pA$ collisions if target $A$ taking from heavy nucleus to proton  (see Fig.11 in [30]). Neutron star provides a completely new energy range for observing the gluon condensation effect since $A^*\gg A$.

\begin{table}[htb!]
\begin{center}\label{tab:1}
\caption{The parameters of the GCE spectra in the gluon condensation mechanism.}
\vspace{1em}
\includegraphics[width=1.0\textwidth]{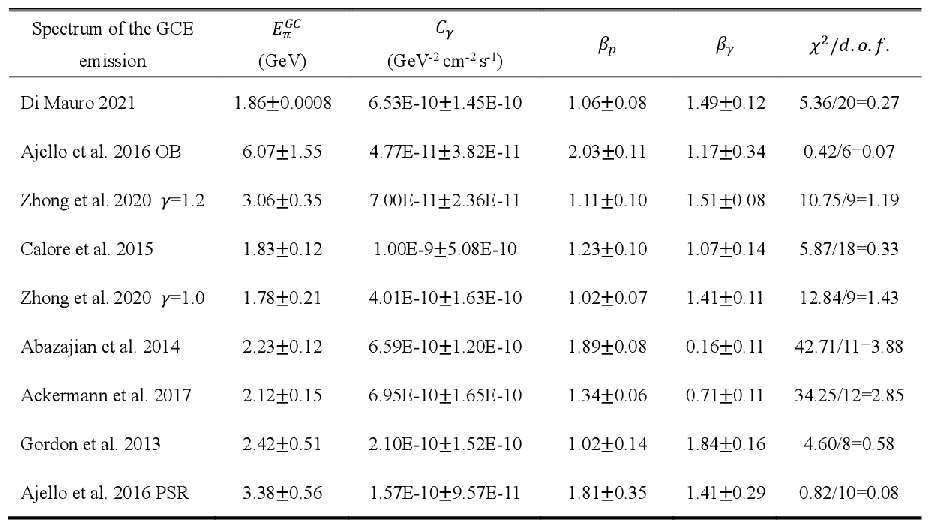}
\vspace{-3em}
\end{center}
\end{table}

    Table 1 shows that $E_{\pi}^{GC}\sim 1~{\rm GeV}$, which are consistent with that in the millisecond pulsars [29].
It is conceivable that the surface of a neutron star has a lattice structure formed by heavy atoms, while the density of neutrons is significantly higher than that of atoms.
The superfluid of neutrons inside the interior can penetrate the lattice space since there is sufficient empty space among the atoms, resulting in the formation of neutron cluster $A^*$ on the star surface.
We assume that these neutron clusters $A^*$ are stable.
Due to these lattices having a similar size, the gluon condensation threshold $E_{\pi}^{GC}$ are restricted to a similar low energy scale.
Therefore, we consider that the GCE originates from the collisions of VHE protons in the Galactic center with the neutron cluster $A^*$ on the surface of neutron stars.

    An important characteristic of the gluon condensation mechanism is that its spectrum presents a straight line in a
double logarithmic coordinator within an interval $[E_{\pi}^{GC}, E_{\pi}^{cut}]$.
Besides, the length of the linear interval increases with $(E_{\pi}^{GC})^2$.
We compare it with a dark matter annihilation model [31].
The shape of the dark matter annihilation spectrum is usually parameterized by a power law with an exponential cutoff
$\Phi=K(E_{\gamma}/E_0)^{-\Gamma}\exp(-E_{\gamma}/E_{cut})$, which is different from the gluon condensation mechanism.
The difference between the two can be clearly seen in Fig.3.

\begin{figure}
  \begin{center}
   \includegraphics[width=0.8\textwidth]{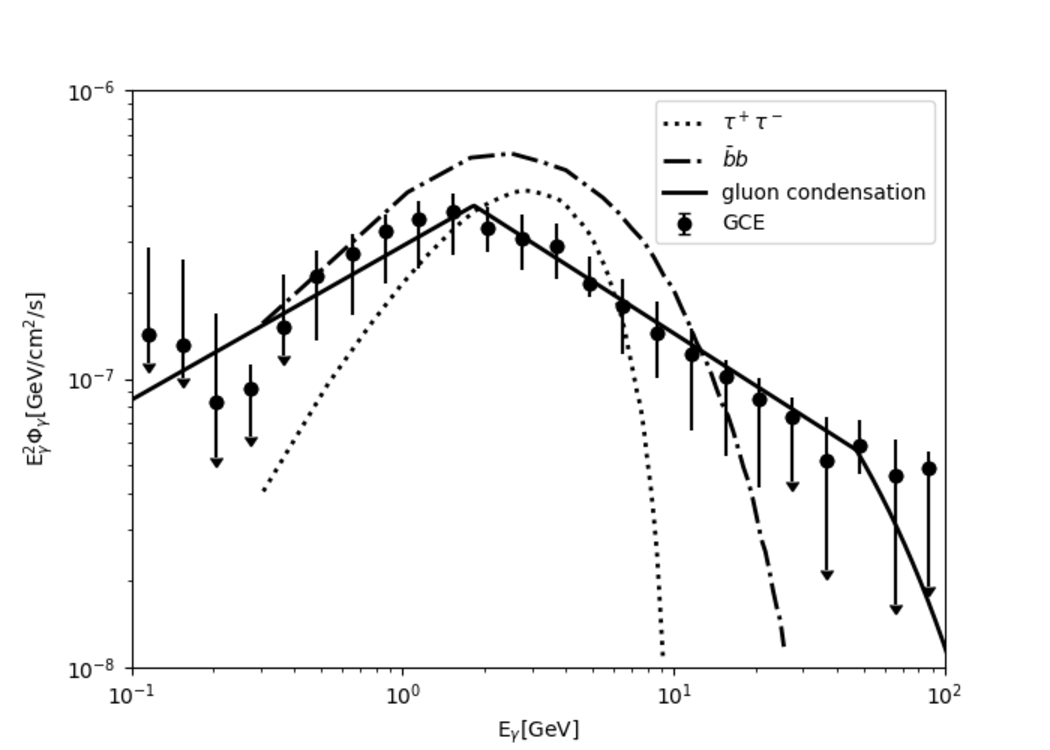}
   \caption{A comparison of the GCE spectra between the gluon condensation mechanism with the dark matter model, the latter is taken from [31] and the curve heights have been adjusted.
}\label{fig:3}
  \end{center}
\end{figure}

\section{Non-rotating neutron stars}

    The measured spectra shown in Fig.1 have some differences between each other since they are derived by
different analysis processes. However, each of these spectra is a synthesis of the sub-spectra of many sources. Curiously each of them exhibits the typical BPL form. We know that it is mathematically almost impossible to synthesize several different BPL curves into one simple BPL. This requires us to make new considerations about the sources of the spectra in Fig. l.

    Using the bright low-mass X-ray binaries population to estimate the number of millisecond pulsars, Cholis, Hooper and Linden
has found millisecond pulsars is not enough to explain the whole excess [32]. However it is also highly depended on the luminosity function of millisecond pulsars.
Dinsmore and Slatyer have examined the luminosity function of millisecond pulsars to explain GCE and found the needed number of millisecond pulsars should be in the range of $\mathcal{O}(10^{4-5})$ [14].
Another possible suggestion that the GCE originates from currently undetected neutron stars with the similar spatial distribution of millisecond pulsars.
The non-rotating (or slower rotating) neutron star is hard to be detected which makes it a possible candidate.
In this work, we attempt to use the gluon condensation effect of the non-rotating (or slower rotating) neutron stars to explain the GCE.

   The Galactic center is a fascinating and mysterious place, filled with high-density materials
such as neutron stars and black holes. When a giant star burns up its nuclear fuel, its core collapses into a very compact sphere, causing its rotation speed to increase rapidly due to the angular momentum conservation. These rotating stars emit pulsed electromagnetic waves that have been identified as pulsars.
Theoretically, not all neutron stars are the pulsars. At the beginning of the formation of a neutron star, its rotation energy may be seriously lost due to other reasons, for say, abandonment of a large number of neutron materials, forming an non-rotating neutron star.

     Non-rotating neutron stars do not have characteristic pulse electromagnetic radiation, making them difficult
to distinguish from other dark stars. Nevertheless, the theoretical study of non-rotating neutron stars is still an interesting topic.
Let us give a few examples. As early 1941, Cowling provided the first classification of modes according to the physics dominating their behaviour [33].
Gittins and Andersson discussed the $r$-modes of slowly rotating, stratified neutron stars [34].
The mass and radius of non-rotating neutron star with maximum-mass
play a crucial role in constraining the elusive equation of state of cold dense matter and in predicting the fate of remnants from binary neutron star mergers [35].

    We only focus on the contribution of the non-rotating neutron stars to the GCE and hope it can provide indirect evidence
of existing non-rotating neutron
stars in the Galactic center. The spatial morphology of the GCE requires the existence of a population of almost spherically symmetric
distribution of non-rotating neutron stars at the Galactic center. We image that neutron stars are formed by the collapse of massive stars, in which high and low rotational (the latter including non-rotating) neutron stars are produced randomly, mainly determined by whether the stars have large primordial angular momentum.
If these non-rotating neutron stars are old neutron stars, produced in early stars and not engulfed by black holes, they are pulled by the gravity of a central black hole and form a three-dimensional symmetric spatial distribution.
If these non-rotating neutron stars are old neutron stars and originate from early stars, they would have the similar spatial distribution of millisecond pulsars which is approximately symmetric spatial distribution.
We emphasize that since there is no direct detection of the non-rotating neutron stars, the above mentioned origin and spatial morphology is an assumption.

    When the VHE protons in the Galactic center collide with these non-rotating neutron stars, they generate gamma
rays with a BPL shape, as predicted by the gluon condensation mechanism.
The $pA^*$ collisions generate spectra peaked around $\sim~1 ~{\rm GeV}$. However, non-rotating neutron star loses its rotational acceleration, and the high-energy protons are not accelerated by the star's rotating electromagnetic fields. Most of these protons are cosmic-ray protons that have been further accelerated by known or unknown mechanisms in the Galactic center. Since these accelerators are shared by non-rotating neutron stars within the observation area of each set in Fig. 1, the colliding protons have the same global power index $\beta_p$ in Eq.(8). Similarly, where the parameter $\beta_{\gamma}$ describes photon loss during flight in a considerable measurement area of the Galactic center, and it is also a constant for the same reason.
The gluon condensation spectra in Eq.(8) has only four parameters, and the remaining parameter $C_{\gamma}$ is determined by the strength of the gamma rays. Consequently, we can observe a total spectrum with a broken power law as shown in Fig.1.

\section{Discussions and summary}

    Why we have not considered the contributions of the "$\pi$-bump" in the traditional hadronic mechanism at
$E_{\gamma}\sim 1~{\rm GeV}$ ? In normal hadronic collisions, a significant portion of the proton's kinetic energy is used to heat secondary particles without the gluon condensation effect. However, with the gluon condensation effect, almost all of the available energy at the C.M. frame is utilized to create prions and follow gamma rays. Therefore, the traditional "$\pi$-bump" near $E_{\gamma}\sim 1~{\rm GeV}$ without the gluon condensation effect is covered by the GC spectra in Fig. 1.

     We estimate the contributions of the gluon condensation mechanism to the Milky Way's diffuse emission along the Galactic
plane if considering the VHE cosmic protons colliding with molecular hydrogen of the interstellar medium.
the gluon condensation threshold $E_{\pi}^{GC}\sim
10~{\rm TeV}$ for the $pp$ or $p$-light nuclei collisions, which is determined by the value of $x_c$
[30]. Using Eqs, (9) and (10), one can find that it requests $E_p^{GC}>10^4~{\rm PeV}$, which exceeds the limit of the proton acceleration mechanism in our Milky Way.

       In summary, the intrinsic GCE spectra present the BPL shape. We propose
a novel interpretation on the origin of the GCE, focusing on the
observed spectral shape. Protons are accelerated at the Galactic center and collide with the neutron cluster on the surface of the non-rotating neutron stars.
Due to the gluon condensation in nucleons, these collisions produce a large number of mesons, which have reached the saturation state and
subsequently generate the BPL in the gamma ray spectra.

{\bf ACKNOWLEDGMENTS}

This work is supported by the National Natural Science Foundation of China (NNSFC) No.11851303 and No.12003069, the National Key Research and Development Program of China No. 2022YFF0503304.

\end{document}